\documentclass{aastex}
\usepackage{emulateapj5}
\usepackage{onecolfloat}

\newcommand{\PNM}{P(N|M)}
\renewcommand{\ng}{\bar{n}_g}
\newcommand{\hvol}{h^{3}{\mathrm{Mpc}}^{-3}}
\newcommand{\hmpc}{h^{-1}\mathrm{Mpc}}
\newcommand{\hkpc}{h^{-1}\mathrm{kpc}}
\newcommand{\hMsun}{h^{-1}M_{\odot}}
\newcommand{\Msun}{M_{\odot}}

\newcommand{\Omegam}{\Omega_{m}}
\newcommand{\Omegab}{\Omega_{b}}
\newcommand{\Omegal}{\Omega_{\Lambda}}

\newcommand{\Mbmin}{M_{b,\mathrm{min}}}
\newcommand{\band}[2]{\ensuremath{^{#1}\!{#2}}}
\newcommand{\Mr}{M_{\band{0.1}{r}}}
\newcommand{\gr}{\band{0.1}{(g-r)}}
\newcommand{\dgin}{\delta_{g,\mathrm{in}}}

\bibpunct[,]{(}{)}{;}{a}{}{,}
\begin{document}

\twocolumn[
\title{Interpreting the relationship between galaxy luminosity, color and 
environment}

\author{
Andreas A. Berlind, \altaffilmark{1}
Michael R. Blanton, \altaffilmark{1}
David W. Hogg, \altaffilmark{1}
David H. Weinberg, \altaffilmark{2}
Romeel Dav\'e, \altaffilmark{3}
Daniel J. Eisenstein \altaffilmark{3}
Neal Katz \altaffilmark{4}
}     

\begin{abstract}
We study the relationship between galaxy luminosity, color, and environment
in a cosmological simulation of galaxy formation.  We compare the predicted
relationship with that found for SDSS galaxies and find that the model
successfully predicts most of the qualitative features seen in the data, but 
also shows some interesting differences.  Specifically, the simulation
predicts that the local density around bright red galaxies is a strong 
increasing function of luminosity, but does not depend much on color at fixed 
luminosity.  Moreover, we show that this is due to central galaxies in dark 
matter halos whose baryonic masses correlate strongly with halo mass.  
The simulation also predicts that the local density around blue galaxies is 
a strong increasing function of color, but does not depend much on luminosity 
at fixed color.  We show that this is due to satellite galaxies in halos 
whose stellar ages correlate with halo mass.  Finally, the simulation fails to 
predict the luminosity dependence of environment observed around low 
luminosity red galaxies.  However, we show that this is most likely due to
the simulation's limited resolution.  A study of a higher resolution, smaller
volume simulation suggests that this dependence is caused by the fact that 
all low luminosity red galaxies are satellites in massive halos, whereas 
intermediate luminosity red galaxies are a mixture of satellites in massive
halos and central galaxies in less massive halos.
\end{abstract}

]

\altaffiltext{1}{Center for Cosmology and Particle Physics, New York 
University, New York, NY 10003, USA; aberlind@cosmo.nyu.edu, 
michael.blanton@nyu.edu, david.hogg@nyu.edu}
\altaffiltext{2}{Department of Astronomy, The Ohio State University, Columbus, 
OH 43210, USA; dhw@astronomy.ohio-state.edu}
\altaffiltext{3}{Steward Observatory, University of Arizona, Tucson, AZ 85721, 
USA; rad@as.arizona.edu, deisenstein@as.arizona.edu}
\altaffiltext{4}{Department of Physics and Astronomy, University of 
Massachusetts, Amherst, MA 01003, USA; nsk@kaka.phast.umass.edu}


\section{Introduction} \label{intro}

One of the keys to understanding galaxy formation and evolution lies in
understanding the connection between the observable properties of galaxies 
and the larger scale environments in which they live.  High signal-to-noise 
correlations between galaxy properties and environment have now been observed 
in many different galaxy surveys of the local universe 
(\citealt{hogg03a,blanton04,gomez03,balogh03,goto03,lewis02,kauffmann04}, and 
references therein).
The biggest challenge (at least at low redshift) thus lies in predicting 
these correlations theoretically and understanding their physical origin.

Using data from the Sloan Digital Sky Survey (SDSS; \citealt{york00}), 
\citet{blanton03b} showed that luminosity and color are the galaxy properties 
most strongly correlated with environment (defined as the overdensity of 
galaxies on a scale of $1\hmpc$ around each galaxy), with other properties 
such as surface brightness or morphology correlating with environment mainly 
secondarily through their correlation with luminosity and color.  
\citet{hogg03a} and \citet{blanton04} explored the dependence of 
environment on these two galaxy properties in detail.  They found that for blue
galaxies, the environment does not depend on luminosity, but that local
density increases with redder color, whereas for red galaxies there is a 
strong relation between luminosity and local density\footnote{\cite{hogg03a} 
and \cite{blanton04} measure the mean environment for fixed galaxy properties, 
instead of the reverse, because environment measures have much lower 
signal-to-noise ratios than measures of galaxy properties.  Averaging galaxy 
properties at fixed local density estimate therefore yields results that are
lower signal-to-noise overall, and much more prone to statistical bias, than
does averaging local density estimates at fixed galaxy properties.
}.  Both very bright ($\Mr\lesssim -22$) and faint ($\Mr\gtrsim -19$) red 
galaxies reside predominantly in highly overdense regions, while galaxies at 
intermediate luminosities reside in less overdense environments.  These 
effects have also been seen in the dependence of the galaxy autocorrelation 
function on color and luminosity (\citealt{norberg02}, I. Zehavi et al. in 
preparation).

In contemporary theories of galaxy formation and evolution, all galaxies reside
in virialized dark matter halos, whether they are field galaxies that occupy
their own halo or group and cluster galaxies that occupy a larger mass halo
along with other galaxies.  The observed properties of galaxies, such as 
color and luminosity, depend on the mass and assembly history of their dark 
matter halos.  The dependence of these properties on environment thus comes 
from their correlation with halo mass, as well as the correlation of halo mass 
and history with the larger scale local mass density.  The latter has been 
studied in N-body simulations and is relatively well understood 
(e.g., \citealt{lemson99}).  It is convenient to think of the correlation
of luminosity and color with environment in these terms, since the problem is 
reduced to understanding the dependence of luminosity and color on halo mass 
and formation history.  This is the route taken by semi-analytic galaxy 
formation models (e.g., \citealt{kauffmann99,benson00}) and hydrodynamic
simulations have recently been analyzed in these terms (\citealt{berlind03}, 
Z. Zheng et al. in preparation).

In this paper we compare the observed \citet{hogg03a} environment vs.
luminosity and color relation with that predicted by a cosmological
hydrodynamic simulation.  Furthermore, we interpret the simulation results
in terms of the dark matter halos that galaxies occupy.


\section{Simulation} \label{models}

We use a smoothed particle hydrodynamics (SPH) simulation of a $\Lambda$CDM 
cosmological model, with $\Omegam=0.4$, $\Omegal=0.6$, $\Omegab=0.02h^{-2}$,
$h\equiv H_0/(100~\mathrm{km~s}^{-1}~\mathrm{Mpc}^{-1})=0.65$, $n=0.95$, and 
$\sigma_8=0.8$.  The simulation uses the Parallel TreeSPH code 
\citep{hernquist89,katz96,dave97} to follow the evolution of $144^3$ gas and 
$144^3$ dark matter particles in a $50\hmpc$ box from $z=49$ to $z=0$.  The 
mass of each dark matter particle is $6.3\times10^9 \Msun$, the mass of 
each baryonic particle is $8.5\times10^8 \Msun$, and the gravitational 
force softening is $\epsilon_{\rm grav}=7\hkpc$ (Plummer equivalent).

Dark matter particles are only affected by gravity, whereas gas particles are 
subject to pressure gradients and shocks, in addition to gravitational forces.
The TreeSPH code includes the effects of both radiative and Compton cooling. 
Star formation is assumed to happen in regions that are Jeans unstable and 
where the gas density is greater than a threshold value 
($n_H \geq 0.1 {\rm cm}^{-3}$) and colder than a threshold temperature 
($T \leq 30,000\,$K).  Once gas is eligible to form stars, it does so at a 
rate proportional to $\rho_{\mathrm{gas}}/t_{\mathrm{gas}}$, where 
$\rho_{\mathrm{gas}}$ is the gas density and $t_{\mathrm{gas}}$ is the longer 
of the gas cooling and dynamical times.  Gas that turns into stars becomes 
collisionless and releases energy back into the surrounding gas via supernova 
explosions.  A Miller-Scalo (\citeyear{miller79}) initial mass function of 
stars is assumed, and stars of mass greater than $8\Msun$ become 
supernovae and inject $10^{51}$ergs of pure thermal energy into neighboring 
gas particles.  The star formation and feedback algorithms are discussed 
extensively by \cite{katz96}, and the particular simulation employed here is
described in greater detail by \cite{murali02}, \cite{dave02}, and
\cite{weinberg04}.  The parameters are all chosen on the basis of 
{\it a priori} theoretical and numerical considerations and are not adjusted 
to match any observations.

SPH galaxies are identified at the sites of local baryonic density maxima using
the SKID algorithm,\footnote{See 
{\tt http://www-hpcc.astro.washington.edu/tools/skid.html} and \cite{katz96}.} 
which selects gravitationally bound groups of star and cold, dense gas 
particles.  Because dissipation greatly increases the density contrast of these
baryonic components (see e.g., \citealt{weinberg04}, Fig.~1), there is 
essentially no ambiguity in the identification of
galaxies.  We retain only those particle groups whose mass exceeds a threshold 
$\Mbmin=5.42\times10^{10} \Msun$, corresponding to the mass of 64 SPH 
particles, and the resulting galaxy space density is $\ng=0.02\hvol$.  This
space density corresponds to galaxies brighter than $\Mr<-18.4$ when
integrating the \citet{blanton03a} SDSS luminosity function (evolved to
redshift zero), corresponding to $L/L^*=0.18$. The galaxy properties that we 
use in this analysis, aside from position and velocity, are the total baryonic 
mass and the median stellar age (i.e., the look-back time to the point at 
which half of the stellar mass had formed).  

To study the effects of numerical resolution, we also analyze a smaller volume,
higher resolution simulation having the same cosmological model.  This 
simulation has $128^3$ gas and $128^3$ dark matter particles in a $22.222\hmpc$
box, yielding particle masses of $1.05\times10^8 \Msun$ and 
$8.8\times10^8 \Msun$ for baryonic and dark matter particles, respectively.
The gravitational force softening scale is $3.5\hkpc$ (Plummer equivalent).
This simulation has eight times higher mass resolution than our main $50\hmpc$ 
simulation, and the lowest mass galaxies that we consider contain 512 baryonic 
particles instead of 64.  In addition to having higher resolution, this 
simulation includes an additional piece of physics: a photoionizing UV 
background (calculated by \citealt{haardt96}).


\section{Comparison to SDSS} \label{comparison}

\citet{hogg03a} showed the dependence of environment on galaxy colors and
luminosities.  In order to compare these results to the predictions of our
SPH simulation, we must measure the same quantities in the simulation that
were measured from the SDSS data.  In \citet{hogg03a}, two different 
definitions of environment are used.  The first, which we call $\delta_{g,1}$,
is a deprojected angular correlation function that recovers the real-space
galaxy density contrast around each galaxy in a spherical Gaussian filter 
$e^{-r^2/2a^2}$ of radius $a=1 \hmpc$ \citep{eisenstein03}.  The second, 
which we call $\delta_{g,8}$, is a simple redshift-space galaxy density 
contrast in a spherical top-hat filter of radius $8 \hmpc$.  In both cases, 
the central galaxy is not counted in the density measurement.  

We approximately reproduce these measurements of environment in the SPH 
simulation.  In order to measure $\delta_{g,1}$ for each SPH galaxy, we 
compute the real-space density of its neighboring galaxies in a spherical 
Gaussian filter of radius $1 \hmpc$.  We then divide that density by 
$\ng=0.02\hvol$, the mean density of galaxies in the simulation cube, and 
subtract one to obtain the density contrast.  In order to measure 
$\delta_{g,8}$, we first put the SPH galaxies in redshift space assuming that 
the line of sight direction is along the $z-$axis of the simulation cube.  We 
then compute the density of neighboring galaxies in a top-hat sphere of radius 
$8 \hmpc$ and convert to density contrast as before.  We also compute 
$\delta_{m,1}$ and $\delta_{m,8}$, the density contrast of dark matter around 
each SPH galaxy using $1\hmpc$ Gaussian and $8\hmpc$ top-hat filters, 
respectively.

\begin{figure}[t]
\epsscale{0.8}
\plotone{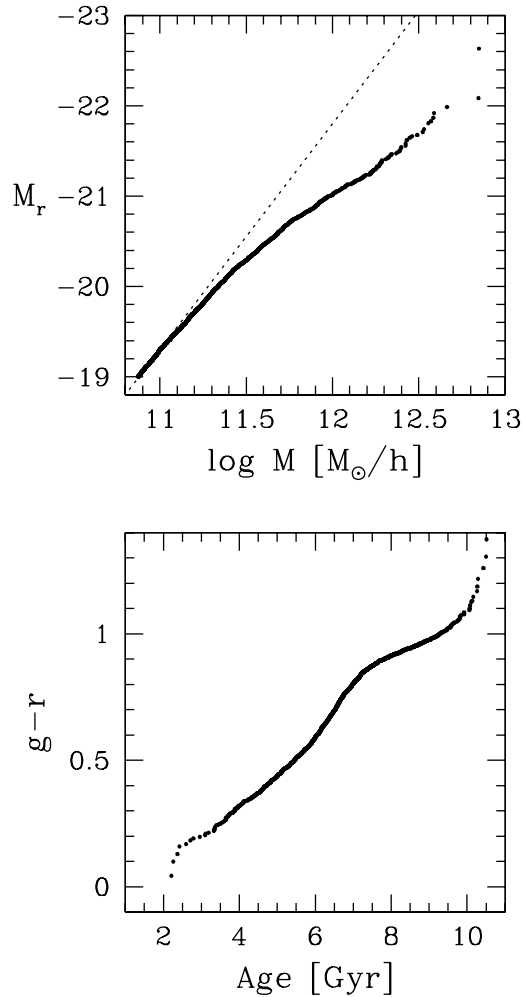}
\caption{
{\it Top panel}:
Monotonic transformation required to convert the baryonic masses of 
SPH galaxies to absolute $\band{0.1}{r}-$band magnitudes so that their 
distribution matches that of SDSS galaxies.  The dotted line shows the linear 
relation between luminosity and baryonic mass.
{\it Bottom panel}:
Monotonic transformation required to convert the median stellar ages
of SPH galaxies to $\gr$ colors so that their distribution matches that of 
SDSS galaxies.
} 
\label{fig:transform}
\end{figure}

We have not computed spectral energy distributions for galaxies in the SPH 
simulation and thus do not predict luminosities and colors.  We therefore use 
the total baryonic mass of each SPH galaxy as a proxy for its luminosity, and 
the median stellar age as a proxy for its color.  Specifically, we assume that 
a galaxy's $\band{0.1}{r}-$band luminosity is a monotonic function of its 
baryonic mass and that its $\gr$ color is a monotonic function of its stellar 
age, and we find the nonlinear transformations that map the simulated mass and 
age distributions to the observed \citet{blanton03b} SDSS $\Mr$ and $\gr$ 
distributions.  These transformations are shown in Figure~\ref{fig:transform}.
The $\Mr$ magnitudes and $\gr$ colors of SPH galaxies that we use henceforth 
are derived in this way.  We could in principle use the star formation 
histories of the simulated galaxies to compute $\Mr$ and $\gr$ directly, but 
since the modeling of star formation is crude and affected by finite 
resolution, we believe that this monotonic mapping procedure is more realistic 
and allows more accurate interpretation of the observed 
luminosity-color-environment relation.

Figure~\ref{fig:transform} shows that the transformation required to 
convert masses to luminosities is strongly nonlinear and significantly 
shallower than the straightforward linear relation shown by the dotted line.
In other words, assuming a roughly constant stellar IMF, the SPH simulation 
produces too many high mass relative to low mass galaxies, yielding a mass 
function that does not cut off steeply enough at the high mass end.  This 
problem has been discussed in previous studies that used the same simulation, 
and it has been partly attributed to numerical resolution effects 
(\citealt{berlind03}; M. Fardal et al., in preparation).  The strange 
shape of the color-age transformation is caused by the fact that the SDSS 
color distribution is bimodal, with a blue peak at $\gr\sim0.5$ and a stronger 
red peak at $\gr\sim0.9$, whereas the SPH age distribution can be described by 
a single Gaussian with a peak at $\sim6.7$ Gyr.

In this study we use the SPH galaxy and mass distributions at a redshift of
zero.  The SDSS galaxies, however, have a median redshift of $z=0.1$.  We have 
checked the SPH predictions at $z=0.125$ and found that there is no significant
evolution out to this small redshift.


\begin{figure*}[t]
\epsscale{1.5}
\plotone{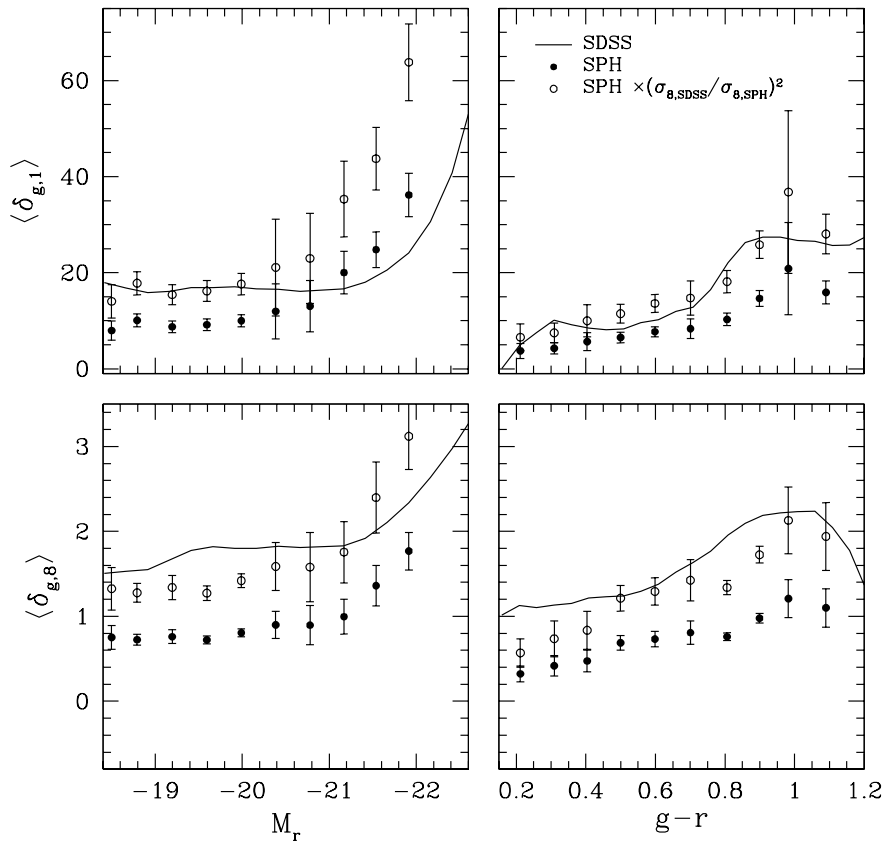}
\caption{
Mean galaxy density contrast $\langle\delta_g\rangle$ measured in bins of 
absolute $\band{0.1}{r}-$band magnitude ({\it left panels}) and $\gr$ color 
({\it right panels}), where $\delta_g$ is measured around galaxies in 
spherical Gaussian filters of radius $1 \hmpc$ ({\it top panels}) and 
spherical top-hat filters of radius $8 \hmpc$ ({\it bottom panels}).
{\it Solid points} show results for SPH galaxies, {\it curves} show the 
\citet{hogg03a} results for SDSS galaxies, and {\it open points} show the SPH
galaxy results multiplied by the ratio of $\sigma_8^2$ for SDSS and SPH 
galaxies.  Absolute $\band{0.1}{r}-$band magnitudes and $\gr$ colors were 
assigned to SPH galaxies according to the transformations shown in 
Fig.~\ref{fig:transform} and discussed in \S~\ref{comparison}.  Error-bars 
show the uncertainty in the mean estimated from the dispersion found among 
the eight octants of the simulation cube.
} 
\label{fig:env1d}
\end{figure*}

\section{Results} \label{results}

Figure~\ref{fig:env1d} shows the dependences of mean environment
($\delta_{g,1}$ and $\delta_{g,8}$) on color and luminosity for both SPH
(solid points) and SDSS (curves) galaxies.  One of the most striking 
differences between the two is that the density contrast around SPH galaxies
is almost always lower than that of SDSS galaxies.  
As discussed by \citet{weinberg04}, the particular realization of initial 
conditions in this $50\hmpc$ cube leads to a low amplitude of the dark matter
correlation function relative to the average of N-body simulations of this
volume.  This statistical fluctuation probably accounts for most of the
difference between the overall level of density contrasts in the simulation
and the data, though the systematic effect of missing power on scales larger
than the box could play a role, and it is possible that the simulation
assumes too low a value of $\sigma_{8,\mathrm{matter}}$ or predicts an 
incorrect value of galaxy bias.  To reduce the impact of the low clustering 
amplitude on our subsequent analysis, we multiply all SPH density contrasts by 
$(\sigma_{8,\mathrm{SDSS}}/\sigma_{8,\mathrm{SPH}})^2 = 1.76$, 
the ratio of clustering amplitudes of SDSS and SPH galaxies on a $8 \hmpc$ 
scale\footnote{We multiply by the square of $\sigma_8$ because our 
measurement is a density contrast of galaxies around galaxies and is thus 
effectively an autocorrelation.}.  This correction brings the SPH density 
contrast levels up to those of the SDSS on both the $1\hmpc$ and 
$8\hmpc$ scales (open points in Figure~\ref{fig:env1d}).

Overall, there is good qualitative agreement between the predicted and observed
environments as a function of luminosity and color.  In both cases, all 
galaxies on average live in overdense regions.  Moreover, the SPH simulation 
successfully predicts that the density contrast around galaxies is independent 
of luminosity at low luminosities, but increases strongly with luminosity for 
brighter galaxies.  The SPH simulation likewise captures the basic dependence 
of environment on color: there is a slow rise of density contrast around 
galaxies as they get redder, followed by a levelling off for the reddest 
galaxies.  

However, Figure~\ref{fig:env1d} also reveals interesting differences between 
theory and observation, the most notable being that the $1\hmpc$ SPH density 
contrast upturn occurs at roughly one magnitude fainter than it does in the 
SDSS.  This most likely indicates that the simple transformation from baryonic 
mass to luminosity shown in Figure~\ref{fig:transform} is not able to repair 
the overproduction of massive galaxies in the SPH simulation.  In order to 
obtain the correct luminosity function, the nonlinear transformation pulls 
many high mass galaxies to low luminosities (relative to the luminosities they 
would be assigned under a linear transformation), but their environments 
remain very dense, resulting in the density upturn shifting to lower 
luminosity.  Adopting a simple linear transformation between mass and 
luminosity would result in the predicted density upturn occurring at the right 
luminosity, but the luminosity function would then be incorrect.  We thus 
conclude that the simulation is successful in predicting the qualitative 
behavior of the luminosity-density relationship, but fails to get the 
luminosity scale right.  It is possible that this discrepancy is an artifact 
of the finite simulation volume, which contains only two clusters of mass
$M \ga 3\times 10^{14}M_\odot$, but it appears significant relative to the 
error bars in Figure 2, which are estimated from the dispersion among the 
eight octants of the cube divided by $8^{1/2}$ to yield the standard error on 
the mean.  No monotonic transformation between stellar mass and luminosity 
would simultaneously reproduce the observed luminosity function and the 
luminosity scale of the density upturn, so a more complicated environment 
dependent error in the stellar masses is needed to fix the discrepancy if it 
is real.  We have tried simply adding a 0.3-magnitude scatter to the 
mass-luminosity relation in Figure~\ref{fig:env1d}, but this does not move the 
upturn to significantly lower luminosity.

\begin{figure*}[tp]
\epsscale{1.7}
\plotone{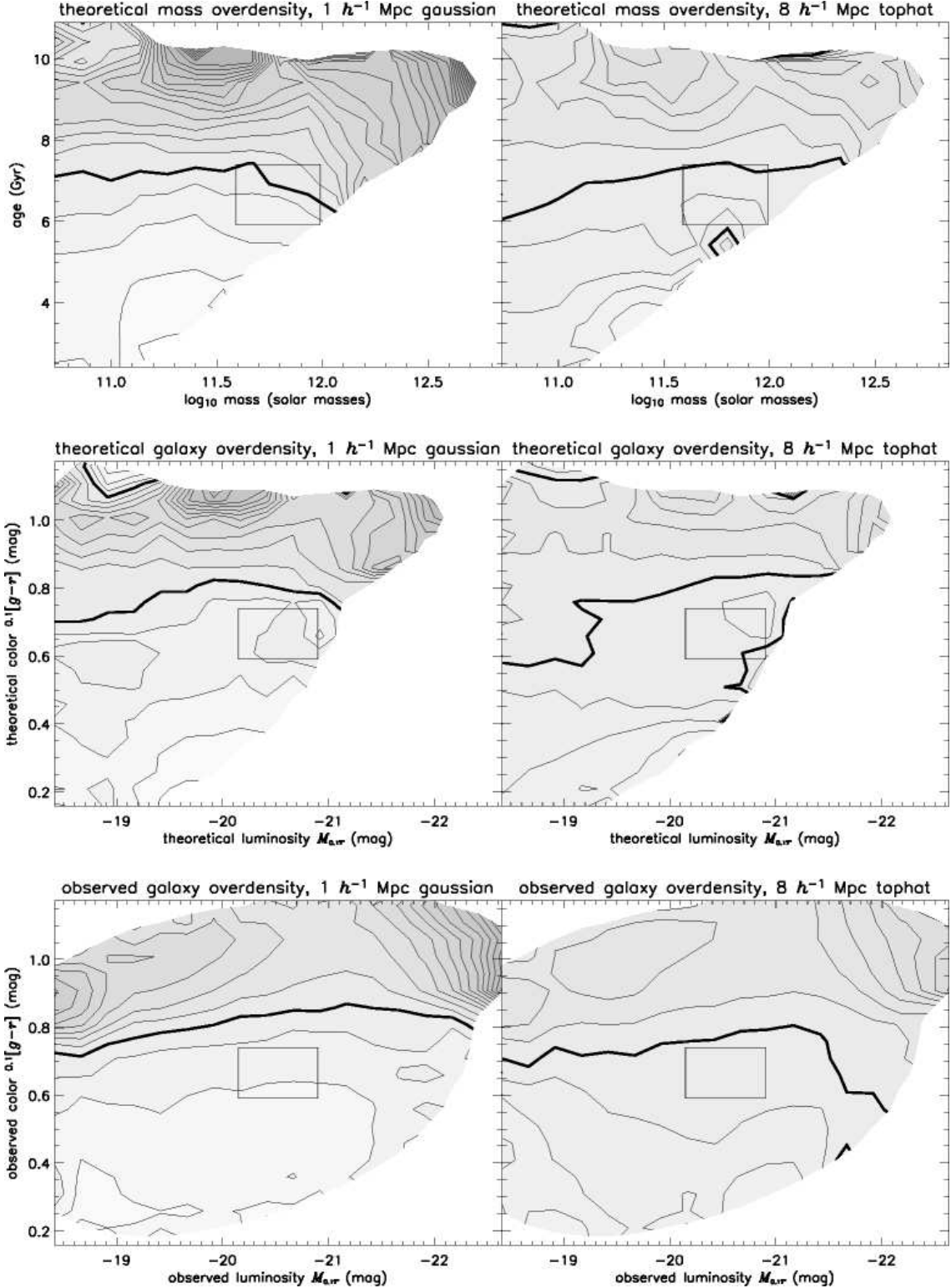}
\caption{
{\it Top panels}:
Mean dark matter density contrast $\langle\delta_m\rangle$ around SPH galaxies 
measured in sliding rectangular bins of galaxy baryonic mass and median stellar
age.  The size and shape of the bins are shown in each panel as a rectangular 
box.
{\it Middle panels}:
Mean galaxy density contrast $\langle\delta_g\rangle$ around SPH galaxies 
measured in sliding rectangular bins of absolute $\band{0.1}{r}-$band magnitude
and $\gr$ color.  Absolute $\band{0.1}{r}-$band magnitudes and $\gr$ colors 
were assigned to SPH galaxies according to the transformations shown in 
Fig.~\ref{fig:transform} and discussed in \S~\ref{comparison}.
{\it Bottom panels}:
$1/V_{max}$-weighted mean galaxy density contrast $\langle\delta_g\rangle$
around SDSS galaxies measured in sliding rectangular bins of absolute 
$\band{0.1}{r}-$band magnitude and $\gr$ color.
{\it All panels}:
Left-hand-side panels show densities computed in spherical Gaussian filters of 
radius $1 \hmpc$ and right-hand-side panels show densities computed in 
spherical top-hat filters of radius $8 \hmpc$.  In each panel, bold contours 
show the global mean density contrast and other contours show $20\%$ 
increments in density contrast (going from lighter to darker regions).
}
\label{fig:env2d}
\end{figure*}

Figure~\ref{fig:env2d} shows the joint mean dependence of environment on color
and luminosity.  The top panels show the purely theoretical SPH dark matter
density contrast as a function of baryonic mass and stellar age. The middle
panels show the SPH analog to observations: galaxy density contrast as a 
function of absolute $\band{0.1}{r}-$band magnitude and $\gr$ color, where 
$\Mr$ and $\gr$ are obtained from the transformations discussed in 
\S~\ref{comparison}.  Finally, the bottom panels show the observed SDSS galaxy 
density contrast as a function of $\Mr$ and $\gr$ (first shown by 
\citealt{hogg03a}).  Overall there is remarkable agreement between theory and 
observation, as well as a few notable differences.  The main conclusions that 
we draw from this comparison are as follows:

1.
The SPH galaxy density contrast as a function of transformed luminosity and
color qualitatively traces the dark matter density contrast as a function
of baryonic mass and stellar age.  This result is reassuring because it 
suggests that the observed measurement of environment is likely also tracing
the underlying dark matter density.

2.
The SPH simulation successfully predicts that the mean density contrast around
blue galaxies ($\gr\lesssim 0.8-0.9$) increases with increasing $\gr$ and 
shows little dependence on luminosity at fixed color.

3.
The SPH simulation successfully predicts that very luminous red galaxies reside
in highly overdense regions, and that their overdensity increases with 
luminosity.  On the other hand, the simulation underpredicts the luminosity 
at which the transition to high density occurs.  This is the same effect that 
we saw in Figure~\ref{fig:env1d} and discussed above.

4.
The SPH simulation fails to reproduce the observational result that faint red 
galaxies also reside in highly overdense regions (relative to more luminous 
red galaxies).  Instead, it predicts that red galaxies at $\Mr\sim-20$ are in 
highly overdense regions.  We show in the next section that this failure is 
probably an artifact of the simulation's finite resolution.


\section{Discussion} \label{discussion}

In order to understand the physical origin of the environmental dependences
seen in the SPH simulation, it is useful to study them in the context of
galaxies occupying dark matter halos\footnote{Note that by ``halo'' we
mean a virialized structure of overdensity $\rho/\bar{\rho}\sim200$,
which may host a single galaxy, a group of galaxies, or even a rich cluster.}.
Galaxies that live in high mass halos will always have high density environment
measurements simply because their halos contain many other galaxies.  Galaxies 
in low mass halos, however, can only have high density environment measurements
if their halos are close to other halos containing galaxies.  This means that 
host halo mass is strongly correlated with measured galaxy environment, and 
therefore any correlations between galaxy properties and halo mass will also 
show up as correlations with environment.  In order to isolate the halo mass 
dependence of galaxy colors and luminosities in the SPH simulation, we first 
identify dark matter halos using a friends-of-friends algorithm \citep{davis85}
with a linking length of 0.173 times the mean interparticle separation.  We 
then measure $\dgin$ around each galaxy, which is the component of 
$\delta_{g,1}$ that only counts galaxies in the same dark matter halo.  
For a given class of galaxies, the relationship between $\dgin$ and 
halo mass is thus fully described by $\PNM$, the probability distribution that 
a halo of mass $M$ contains $N$ galaxies of this class (see \citealt{berlind02}
for a thorough discussion of this distribution).  This distribution 
was shown for this SPH simulation by \citet{berlind03}, as a function of
galaxy baryonic mass and stellar population age.

\citet{berlind03} also showed that central SPH galaxies within dark matter 
halos form a distinct population from non-central (satellite) galaxies.  In 
particular, they showed that the central galaxy in each halo is almost always 
the most massive and often the oldest galaxy in the halo.  That result suggests
investigating the luminosity-color-environment relation separately for central 
and satellite galaxies.  

\begin{figure}[tp]
\epsscale{0.9}
\plotone{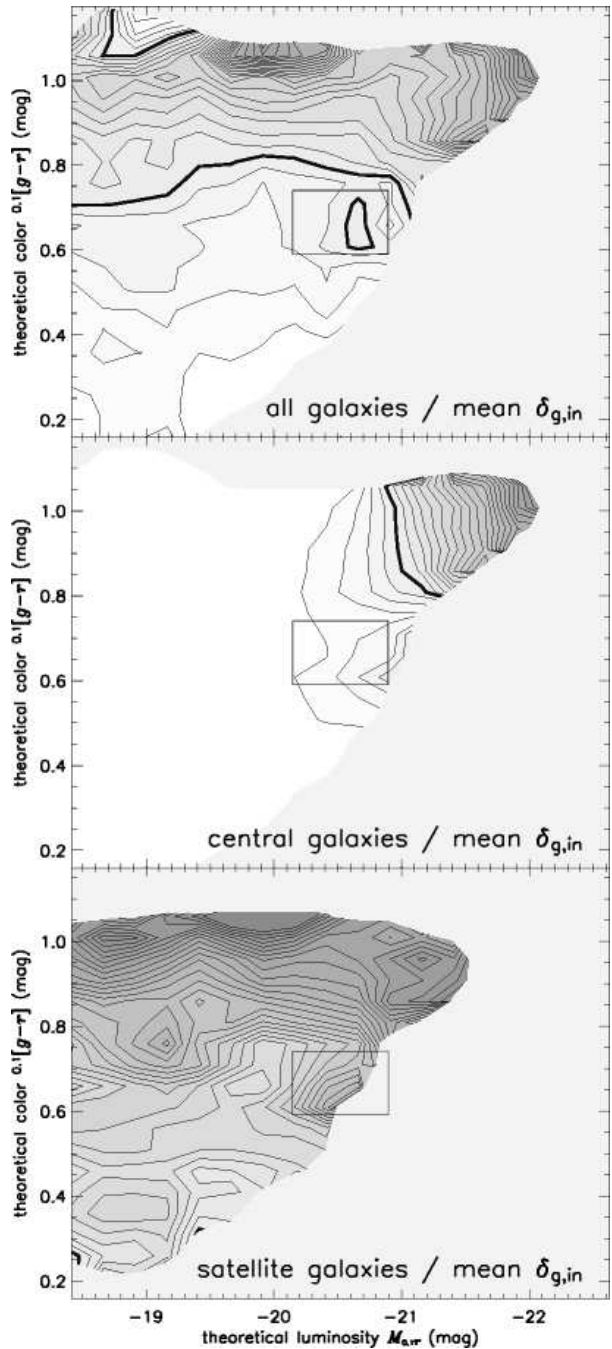}
\caption{
Mean galaxy density contrast $\langle\dgin\rangle$ around SPH galaxies
(computed in spherical Gaussian filters of radius $1 \hmpc$), counting only 
galaxies that are in the same dark matter halos as the SPH galaxies around 
which densities are computed.  This relation is shown for all SPH galaxies 
({\it top panel}), for SPH galaxies that are central in their dark matter 
halos ({\it middle panel}), and for SPH galaxies that are satellites in their 
dark matter halos ({\it bottom panel}).  Other features are same as in 
Fig.~\ref{fig:env2d}.  Note that though there are no density contours at low 
luminosities in the middle panel, there are actually many galaxies in that 
region, but they are mostly isolated single galaxies in halos and thus have a 
very low, uniform $\langle\dgin\rangle$.
}
\label{fig:env2d2}
\end{figure}

These tests are shown in Figure~\ref{fig:env2d2}, which plots the mean
dependence of $\dgin$ on color and luminosity for all (top panel), 
central (middle panel), and satellite (bottom panel) galaxies.  
The striking similarity of the top panel with the middle-left panel of 
Figure~\ref{fig:env2d}, which shows $\delta_{g,1}$ instead of $\dgin$,
demonstrates that the environment on a scale of $1\hmpc$ essentially traces the
population of the galaxy's host dark matter halo --- galaxies in separate
virialized structures make minimal contribution to $\delta_{g,1}$, at least
for purposes of this plot.  This suggests that the observed \citet{hogg03a} 
relation between luminosity, color, and environment is dominated by the 
underlying relation between galaxy luminosity, color, and dark matter halo 
mass.  Any residual relation with the larger-scale density is not as important.

The middle panel in Figure~\ref{fig:env2d2} demonstrates that $\dgin$
(and thus dark matter halo mass) for central galaxies is strongly correlated 
with luminosity (baryonic mass), and mostly uncorrelated with central 
galaxy color (median stellar age) at a fixed luminosity.  It is not surprising
that more massive halos contain more luminous central galaxies, since they have
more baryonic mass, and this relation was shown for this SPH simulation by 
\citet{berlind03} (their Figure~18).  However, it is interesting that there 
is little residual relation between halo mass and central galaxy color.  
This implies that halo mass is much more tightly correlated with central 
galaxy luminosity than with color.  In other words, a fixed mass corresponds 
to a roughly fixed luminosity, but a broad range in color, thus causing 
the contours to be vertical.  The large scatter in color is probably due to 
the fact that central galaxy stellar ages are closely related to the 
formation times of their host dark matter halos, and the distribution of halo 
formation times for a given mass is broad 
(e.g., \citealt{lacey94,wechsler02}).  Overall, to the extent that the
simulation is representative of the real universe, we conclude that the 
strong luminosity dependence and weak color dependence of environment for
bright red galaxies observed by \citet{hogg03a} reflects the strong and weak
dependences of central galaxy luminosity and stellar population age on halo
mass.

The bottom panel in Figure~\ref{fig:env2d2} demonstrates that $\dgin$
for satellite galaxies shows the opposite behavior from that of central 
galaxies: it is strongly correlated with color, and mostly uncorrelated with 
luminosity at fixed color.  Furthermore, comparison to the top panel shows
that the environment dependence for galaxies fainter than $\Mr\sim-21$ tracks 
that of satellite galaxies.  Low luminosity and blue {\it central} galaxies 
live in low mass halos with low $\dgin$, so although they lower the mean 
$\dgin$ by dilution, the trends of $\dgin$ with galaxy properties are driven 
by the satellites in more massive halos with higher $\dgin$.  The luminosity 
of a satellite galaxy need not be strongly correlated with its host halo mass, 
because the galaxy experienced most of its growth while it was the central 
object of a smaller, independent halo.  The strong correlation of satellite 
color with halo mass probably reflects a combination of two effects: (1) an 
earlier onset of star formation in high density regions where halos both 
collapse early and end up merging into massive systems, and (2) the truncation 
of star formation when a galaxy falls into a larger halo and suffers a 
reduction of gas accretion.  Again, to the extent that the simulation is 
representative of the real universe, we can conclude that the strong color 
dependence and weak luminosity dependence of environment for intermediate 
luminosity, blue galaxies found by \citet{hogg03a} reflects these trends for 
satellite galaxies in halos.

\begin{figure}[tp]
\epsscale{1}
\plotone{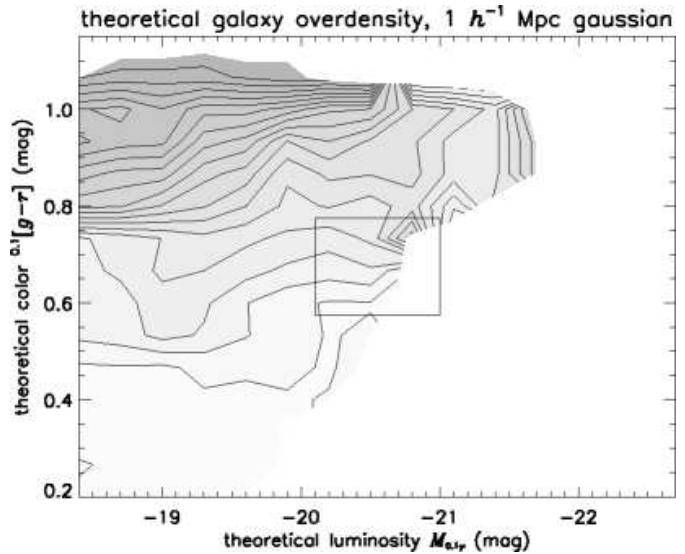}
\caption{
Similar to the middle-left panel of Fig.~\ref{fig:env2d}, except that the 
simulation used is a higher resolution, smaller volume simulation, which is 
described in \S~\ref{models}.
}
\label{fig:env2d_22}
\end{figure}

While the success of the SPH simulation in reproducing these qualitative 
trends is encouraging, it is equally interesting that the SPH simulation fails 
to reproduce the observed environmental dependence for faint red galaxies.  
Low luminosity galaxies are less well resolved by the simulation and their 
star formation histories are less accurately modeled; in particular, galaxies
close to our adopted $64 m_{\mathrm{SPH}}$ resolution threshold have accurate
baryonic masses (stars plus cold gas) but artificially low star formation rates
and stellar masses because their gas densities are systematically 
underestimated (M. Fardal et al., in preparation).  To test whether the 
discrepancy with faint red galaxies is an artifact of limited resolution, we 
analyze the higher resolution, smaller volume simulation described in 
\S~\ref{models}.  Figure~\ref{fig:env2d_22} shows the joint mean dependence of 
galaxy density contrast (defined using a spherical $1\hmpc$ Gaussian filter) as
a function of absolute $\band{0.1}{r}-$band magnitude and $\gr$ color, where 
$\Mr$ and $\gr$ are obtained from the transformations discussed in 
\S~\ref{comparison}.  This figure is identical to the middle-left panel of 
Figure~\ref{fig:env2d}, except that it shows the result for the higher 
resolution simulation.  Due to the smaller volume of this simulation, the 
number of galaxies available is small (there are only 221 galaxies with 
baryonic masses greater than $5.42\times10^{10}\Msun$), yielding poor 
statistics.  The number of galaxies is especially low in the high luminosity 
regime, which is why the environmental dependence for bright red galaxies is 
not obvious in this figure.  Nevertheless, it is clear that low luminosity, 
red galaxies ($\Mr\gtrsim-19$, $\gr\gtrsim0.8$) are in highly overdense 
regions, with higher luminosity red galaxies living in less overdense regions.
The results for low luminosity red galaxies in this simulation are much closer 
to what is seen in the observations (shown in the bottom left panel of 
Fig.~\ref{fig:env2d}).
In contrast, the lower resolution simulation lacks the low luminosity red
feature and only succeeds in reproducing its high luminosity tail at 
$\Mr\sim-20$.

The $22.222\hmpc$ simulation happens to contain one very massive cluster of 
galaxies (whose dark matter halo's viral mass is $\sim 2\times10^{14}\hMsun$).
The low luminosity red galaxies that create the overdense feature discussed
above are essentially all satellite galaxies of this massive cluster.  Low
luminosity central galaxies that occupy low mass halos are never old enough
to populate the same region of Figure~\ref{fig:env2d_22} and thus do not dilute
the high overdensity signal.  As we move to higher luminosities however,
the number of satellite galaxies drops drastically because the halos that are
large enough to house them become more massive than $M_*$ (the characteristic
mass scale in the halo mass function).  Moreover, at these higher luminosities,
isolated central galaxies can be older and thus populate a redder part of the 
color-luminosity space in Figure~\ref{fig:env2d_22}.  They can therefore 
dilute the high overdensity signal coming from the few intermediate luminosity 
red satellite galaxies in massive clusters and hence cause the saddle point 
seen in the figure.  This idea, that central galaxies of low mass halos are 
blue and low luminosity red galaxies must therefore be satellites in more 
massive halos, is independently supported by the work of I. Zehavi et al. (in 
preparation), who model the luminosity and color dependence of the two-point 
correlation function using halo occupation distribution models.

We suspect that spurious scatter in the stellar ages of galaxies near the 
resolution limit of the lower resolution simulation mixes the populations of 
low luminosity central galaxies in low mass halos with low luminosity 
satellites in massive halos and thus erases the low luminosity red galaxy 
feature.  An additional effect of low resolution that could contribute
to this problem is overmerging of low mass satellites in massive halos.  This 
would reduce their numbers and consequently their overdensity measurements.
The higher resolution simulation also differs from the lower resolution one
in that it includes a photoionizing UV background.  It is possible that this
additional piece of physics contributes to the faint red feature seen in 
Figure~\ref{fig:env2d_22}, but not in Figure~\ref{fig:env2d} though it would
be surprising for photoionization to have an important effect at such a high 
mass scale.  Finally, it is possible (though unlikely) that the faint red 
feature seen in the SDSS data is caused by other physical mechanisms that are 
not included in these simulations and that the feature seen in 
Figure~\ref{fig:env2d_22} is a statistical fluke resulting from only one 
unusually red cluster.  Similar predictions by semi-analytic galaxy formation 
models and pure dark matter high resolution simulations should help to resolve 
this issue.

\section{Summary} \label{summary}

In sum, these results suggest a straightforward interpretation of the main 
trends in the environment-luminosity-color relation found by \citet{hogg03a} 
and \citet{blanton04} in the SDSS.  

1.
Measurements of local galaxy density around galaxies essentially trace the 
mass of their underlying dark matter halos.  Contributions from neighboring 
halos are not as important.   

2.
The environment dependence for luminous red galaxies reflects the trends for 
central galaxies in massive (cluster sized) halos.  The environment depends 
strongly on luminosity and weakly on color because for central galaxies, these
quantities are strongly and weakly correlated with halo mass, respectively. 

3.
The environment dependence for lower luminosity red galaxies reflects the 
changing mixture of central and satellite galaxies as a function of color and 
luminosity.  Low luminosity red galaxies are all satellites in high mass halos 
(whose central objects are luminous red galaxies) --- low luminosity central 
galaxies are always bluer and thus do not populate the same part of the color 
luminosity space.  However, intermediate luminosity red galaxies can be both 
satellites in very massive halos and central galaxies in low mass halos.  
They thus reside, on average, in lower mass halos (and consequently in lower 
density regions) than either their higher or lower luminosity counterparts.

4. 
The environment dependence for blue galaxies reflects the trends for 
satellite galaxies in intermediate mass (group sized) halos.  For this 
population, luminosity is only weakly correlated with host halo mass (and thus 
environment), while color is strongly correlated.

\acknowledgments 

We thank James Bullock, Andrey Kravtsov and Risa Wechsler for useful 
discussion and feedback.
This research made use of the NASA Astrophysics Data System.  AAB, MRB and DWH 
are partially supported by NASA (grant NAG5-11669) and NSF (grant PHY-0101738).
DJE is supported by NSF (grant AST-0098577) and by an Alfred P. Sloan Research 
Fellowship. NK is supported by NSF (grant AST-0205969) and NASA 
(grant NAG5-13102).

Funding for the creation and distribution of the SDSS has been provided by the 
Alfred P. Sloan Foundation, the Participating Institutions, NASA, the NSF, the 
U.S. Department of Energy, the Japanese Monbukagakusho, and the Max Planck 
Society.

\end{document}